\documentclass[preprint,aps,%superscriptaddress,
showpacs%,nofootinbib,eqsecnum,preprintnumbers,twocolumn,prl
]{revtex4}
\usepackage{epsfig}
\begin{document}
%
%\draft
\preprint{}
\date{\today
}
\title{
Asymptotic States in Two Black Hole Moduli Space
}
\author{Kenji~Sakamoto}
\email{sakamoto@het.phys.sci.osaka-u.ac.jp}
\affiliation{Department of Physics, \\
Osaka University, \\
Toyonaka, Osaka 560-0043, Japan}

\begin{abstract}
%%%%%%%%%%%%%%%%%%%%%%%%%%%%%%%%%%%%%%%%%%%%%%%%%%%%%%%%%%%%%%%%%%%%%%
We discuss the quantum states in the moduli space, 
which constructed with maximally charged dilaton black holes. 
Considering the quantum mechanics in the moduli space, 
we obtain the asymptotic states for the near-coincident black holes and the widely separated black holes. 
We study the scattering process of the dilaton black holes with the asymptotic states. 
In the scattering process, the quantum effects in the black hole moduli space 
are investigated. 
%%%%%%%%%%%%%%%%%%%%%%%%%%%%%%%%%%%%%%%%%%%%%%%%%%%%%%%%%%%%%%%%%%%%%%
\end{abstract}
\maketitle
%%%%%%%%%%%%%%%%%%%%%%%%%%%%%%%%%%%%%%%%%%%%%%%%%%%%%%%%%%%%%%%%%%%%%%
%%%%%%%%%%%%%%%%%%%%%%%%%%%%%%%%%%%%%%%%%%%%%%%%%%%%%%%%%%%%%%%%%%%%%%
%%%%%%%%%%%%%%%%%%%%%%%%%%%%%%%%%%%%%%%%%%%%%%%%%%%%%%%%%%%%%%%%%%%%%%
\section{Introduction}
%The study of quantum multi-black holes has recently been a great interest. 
In general relativity, the various solutions for multi-black holes have been discussed. 
The solution describing the static maximally charged black holes,
which have charge of the same sign, was discussed 
by Majumdar and Papapetrou \cite{Papapetrou:1947ib, Majumdar:eu}. 
The multi-Schwarzschild solution was given by Israel and Khan \cite{IsraelKhan}. 
In the dilaton coupled Einstein-Maxwell system, the static multi-black hole solution was studied by Shiraishi \cite{KS3}. 
Other multi-black hole solutions had been studied \cite{Myers:rx, Kastor:1992nn}. 
Recently, the exact solutions of Einstein-Maxwell equation that describe the static pairs of oppositely charged extremal black holes (black diholes) have 
attracted much attention \cite{Emparan:1999au,Emparan:2001bb}. 
The s-brane solutions are given by double Wick rotating the black dihole solutions \cite{Jones:2004rg}. 

The quantum theory of multi-black holes has been focused upon. 
The low-lying quantum states are governed by quantum mechanics in the moduli space, which is constructed with multi-black holes. 
Gibbons and Ruback studied the effective theories in moduli space of Reissner-Nordstr\"{o}m multi-black holes \cite{Gibbons}. 
The scattering in moduli space was discussed by Ferrell et al. \cite{fe, TrFe}, where the black hole-black hole scattering and coalescence were considered in the moduli space which is constructed with (3+1) dimensional Reissner-Nordst\"{o}m black holes. 
The moduli space of multi-black holes with dilaton was discussed by Shiraishi \cite{shir, Shiraishi:1992ct}. 
Dilaton black holes are corresponding to the theory derived from the low energy string and M theory compactified down to four dimensions. 
Therefore, the quantum properties of dilaton black holes have attracted much attention. 

In this paper, we consider the asymptotic states in the moduli space consisting of maximally charged dilaton black holes and discuss the scattering process of the dilaton black holes. 
The scattering of maximally charged dilaton black holes was calculated with the WKB approximation \cite{Sakamoto:2003ju}. 
We compare the scattering process considered from asymptotic states with the results of \cite{Sakamoto:2003ju}. 
In section II, we show the moduli space structure of two-black hole system with dilaton coupling. 
In section III, we consider the quantum mechanics in two black hole moduli space. The Schr\"{o}dinger equation in the moduli space is constructed from the moduli space metric. 
From the feature of potential, we discuss the scattering process. 
In section IV, we consider the asymptotic solutions of the Schr\"{o}dinger equation in two black hole moduli space. Then we find that the scattering processes obtained from asymptotic solutions are corresponding to the numerical results in \cite{Sakamoto:2003ju}. 
%%%%%%%%%%%%%%%%%%%%%%%%%%%%%%%%%%%%%%%%%%%%%%%%%%%%%%%%%%%%%%%%%%%%%%%%%%%%
\section{The Moduli Space Metric for the System of Maximally Charged 
Dilaton Black Holes}
In this section, we consider the moduli space of the Einstein-Maxwell-dilaton system along the technique discussed by Ferrell and Eardley \cite{fe}. 
The Einstein-Maxwell-dilaton system contains a dilaton field $\phi$ coupled to a $U(1)$ gauge field $A_{\mu}$ besides the Einstein-Hilbert gravity.
In the $(3+1)$ dimensions, the action for the fields with particle sources is 
\begin{equation}
S=\int d^{4}x \frac{\sqrt{-g}}{16\pi} \left[ R -
2 (\nabla \phi)^{2} - e^{-2a\phi} F^{2} \right]-\sum^n_{A=1} \int ds_A \left( m_A e^{-2a\phi} +Q_A {\bf A}\frac{d{\bf x}_A}{ds_A}\right),\label{eq:action}
\end{equation}
where $R$ is the scalar curvature and $F_{\mu\nu}=\partial_\mu A_\nu-\partial_\nu A_\mu$. $ds_A$ is the line element of the center of the $A$-th black hole. $m_A$ and $Q_A$ are the mass and the electric charge of the $A$-th black hole. We set the Newton constant $G=1$. 
The dilaton coupling constant $a$ can be assumed to be a positive value
\footnote{This system can be expanded into the dilaton supergravity. 
The value of dilaton coupling is related to the amount of 
supersymmetry preserved by the black holes. 
In the case of no dilaton, Maloney et al. have discussed the 
superconformal multi-black hole system \cite{as2}. 
For $a^2=1$, the action (\ref{eq:action}) reduces to the one which is derived from the low-energy string therory. 
In this paper, we consider the bosonic parts.} .

The metric for the $n$-body system of maximally-charged dilaton black holes has been known as \cite{shir}
\begin{equation}
ds^2=-U^{-2}({\bf{x}}) dt^2+U^{2}({\bf{x}})d{\bf{x}}^2, 
\end{equation}
where
\begin{eqnarray}
U({\bf{x}})&=&(F({\bf{x}}))^{1/(1+a^2)}, \label{eq:U} \\
F({\bf{x}})&=&1+\sum^n_{A=1}\frac{\mu_A}{|{\bf{x}}-{\bf{x}}_A|}. \label{eq:F}
\end{eqnarray}

Using these expressions, the vector one form and dilaton configuration are written as
\begin{eqnarray}
A=\sqrt{\frac{1}{1+a^2}}\left(1-\frac{1}{F({\bf{x}})}\right)dt, \\
e^{-2a\phi}=(F({\bf{x}}))^{2a^2/(1+a^2)}.
\end{eqnarray}
In this solution, the asymptotic value of $\phi$ is fixed to be zero.

The electric charge $Q_A$ of each black hole are associated with the corresponding mass $m_A$ by 
\begin{eqnarray}
m_A&=&\frac{1}{1+a^2}\mu_A,\\
|Q_A|&=&\sqrt{\frac{1}{1+a^2}}\mu_A. 
\end{eqnarray}
These relations between mass and electric charge correspond to the condition of extremal black holes. 
In the special case of $a=0$ for $N=3$, this solution becomes the Papapetrou-Majumdar solution \cite{Papapetrou:1947ib, Majumdar:eu}. 

We consider that the perturbed metric and potential can be written by
\begin{eqnarray}
ds^2=-U^{-2}({\bf{x}})dt^2+2{\bf{N}}d{\bf{x}}dt+U({\bf{x}})d{\bf{x}}^2, \\
A=\sqrt{\frac{1}{1+a^2}}(1-\frac{1}{F({\bf{x}})})dt+{\bf{A}}d{\bf{x}}, 
\end{eqnarray}
where $U({\bf{x}})$ and $F({\bf{x}})$ are defined by (\ref{eq:U}) and (\ref{eq:F}). 
We have only to solve linearized equations with perturbed sources up to $O(v)$ 
for $N_i$ and $A_i$. 
(Here $v$ represents the velocity of the black hole as a point source.) 
We should note that each source plays the role of a maximally charged dilaton black hole.

Solving the Einstein-Maxwell equations and substituting the solutions, 
the perturbed dilaton field and sources to the action (\ref{eq:action}) with proper boundary terms, 
we get the effective action up to $O(v^2)$ for $n$-maximally charged dilaton black hole system
\begin{eqnarray}
S&=&\int dt\left(\frac{1}{2}\sum^n_A m_A {\bf v}_A^2 \right) +\frac{3-a^2}{8\pi(1+a^2)^2}\int d^{4}x (F({\bf{x}}))^\frac{2(1-a^2)}{1+a^2} \nonumber \\
& &\hspace{.5cm} 
 \times \sum^n_{A,B=1}\frac{\mu_A\mu_B}{|{\bf{r}}_A|^{3}|{\bf{r}}_B|^{3}}\biggl[\frac{1}{2}({\bf{r}}_A \cdot {\bf{r}}_B)|{\bf{v}}_A-{\bf{v}}_B|^2-({\bf{r}}_A \times {\bf{r}}_B)\cdot({\bf{v}}_A \times {\bf{v}}_B) \biggl] , \label{lag1}
\end{eqnarray}
where ${\bf{r}}_A={\bf{x}}-{\bf{x}}_A$. $F({\bf{x}})$ is defined by (\ref{eq:F}).

In general, a naive integration in equation (\ref{lag1}) diverges. 
Therefore, we regularize the divergent terms proportional 
to $\int d^3x\delta^3(x)/|x|^p \ (p>0)$ 
which appear when the integrand is expanded must be regularized \cite{InPl}. We set them to zero. 

After regularization, the effective action for two body system (consisting of black holes labeled with $a$ and $b$) can be rewritten as
\begin{eqnarray}
S_{2B}=\int dt
\frac{1}{2} \mu {\bf{v}}^2 
\Biggl[ 1&-&\frac{M}{\mu}
-\frac{(3-a^2)M}{r}
+\frac{M}{m_a} \left( 1+(1+a^2)\frac{m_a}{r} \right) ^{(3-a^2)/(1+a^2)} \nonumber \\
& & \hspace{2cm} +\frac{M}{m_b} \left( 1+(1+a^2)\frac{m_b}{r} \right) ^{(3-a^2)/(1+a^2)}
\Biggl], 
\end{eqnarray}
where $M=m_a+m_b$, $\mu=m_am_b/M$, ${\bf{v}}={\bf{v}}_a-{\bf{v}}_b$ and 
$r=|{\bf{x}}_a-{\bf{x}}_b|$.
Thus the metric of the three dimensional moduli space for two-body system is 
\begin{equation}
g_{ij}=\gamma(r)\delta_{ij},\label{mmet1}
\end{equation}
with 
\begin{equation}
\gamma(r)= 1-\frac{M}{\mu}
-\frac{(3-a^2)M}{r}+\frac{M}{m_a} \left( 1+
\frac{(1+a^2)m_a}{r} \right) ^{\frac{3-a^2}{1+a^2}}+\frac{M}{m_b} \left( 1+
\frac{(1+a^2)m_b}{r} \right) ^{\frac{3-a^2}{1+a^2}}. \label{modmet2}
\end{equation}
The moduli space metric (\ref{mmet1}) depends on the dilaton coupling. 
We notice that the behavior of the moduli metric changes at $a^2=1/3$, 
where the moduli space has a critical structure. 
Therefore, the moduli space geometry has the various structure \cite{shir}. 
%%%%%%%%%%%%%%%%%%%%%%%%%%%%%%%%%%%%%%%%%%%%%%%%%%%%%%%%%%%%%%%%%%%%%%
\section{Quantum mechanics in two-black hole moduli space} 
We consider the quantum mechanics in the moduli space. 
The effective theories of quantum mechanics in moduli space of 
Reissner-Nordst\"{o}m black holes were constructed in \cite{TrFe}. 
In the case of two maximally charged black holes with dilaton, we will study the quantum mechanics. 

Let us introduce a wave function $\Psi$ in the moduli space, which obeys the 
Schr\"{o}dinger equation:
\begin{equation}
i\hbar\frac{d\Psi}{dt}=-\frac{\hbar^2}{2\mu}\nabla^2 \Psi, \label{4scheq1}
\end{equation}
where $\nabla^2$ is the covariant Laplacian constructed from the moduli space metric (\ref{mmet1}). 

The partial wave in a stationary state is 
\begin{equation}
\Psi=\psi_{ql}(r)Y_{lm}(\theta,\phi)\exp(-iEt/\hbar), 
\end{equation}
where $Y_{lm}(\theta,\phi)$ is the spherical harmonic function. 
We define the energy and variables as 
\begin{eqnarray}
E&=&\frac{\hbar^2 q^2}{2\mu}, \\
R&=&\int \sqrt{\gamma} dr, \label{varR} \\
\psi&=&\frac{\chi}{r\sqrt{\gamma}}.
\end{eqnarray}
The Schr\"{o}dinger equation (\ref{4scheq1}) is rewritten as 
\begin{equation}    
\frac{d^2\chi}{dR^2}+(q^2-V)\chi=0, \label{4scheq2}
\end{equation}
where the potential $V$ is represented as 
\begin{equation}
V=\frac{\gamma(r\gamma')'-r(\gamma')^2}{2r\gamma^3}+\frac{l(l+1)}{r^2\gamma}. \label{VVpot}
\end{equation}
Here $'$ stands for $\frac{d}{dr}$ and $l$ is the angular momentum. 

The variable $r$ and $R$ are the distance between two black holes. 
The origin $r=0$ of the moduli space represents the point that the two black holes approach to each other infinitely. The limit of $r\to0$ stands for the near-coincident two black holes. 
The limit of $r\to \infty$ is corresponding to the widely separated two black holes. 
From the equation (\ref{varR}), the limit of $r\to 0$ corresponds to $R\to -\infty$ in the case of $a^2=0$ and $1/3$. 
In the case of $a^2=1$, $r\to0$ becomes $R\to0$. 

\begin{figure}[htpb]
\centering
\includegraphics{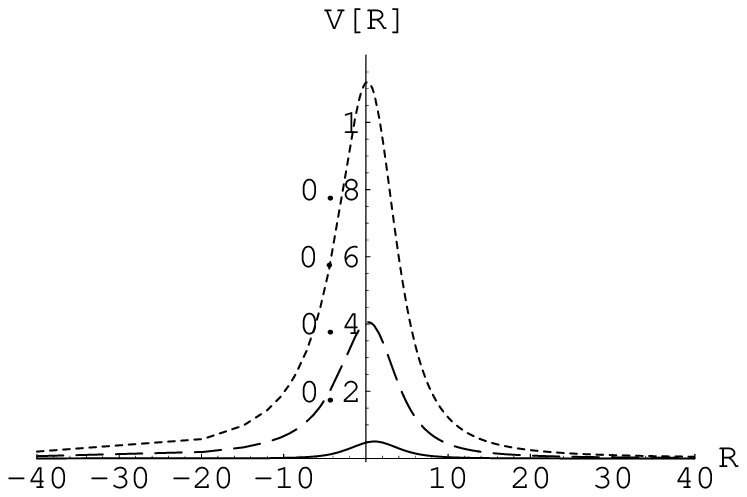}\\
\vspace*{-3mm}
A

\vspace*{10mm}
\includegraphics{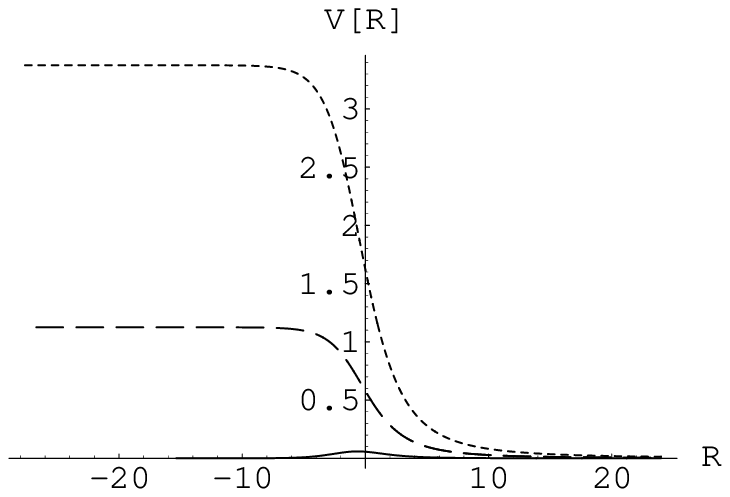}\\
\vspace*{-3mm}
B

\vspace*{10mm}
\includegraphics{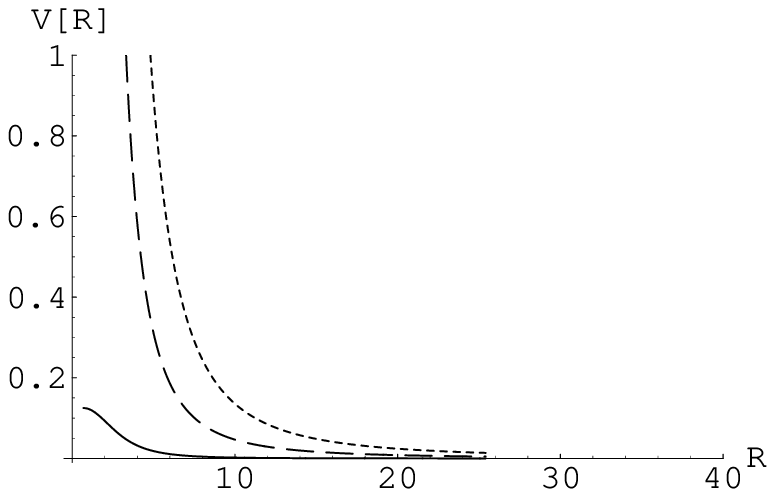}\\
\vspace*{-3mm}
C
\caption{The potential $V$ as the function of $R$ for the angular 
momentum $l=0$ (solid), $l=1$(dashed) and $l=2$(dotted).  
A for $a^2=0$, B for $a^2=1/3$ and C for $a^2=1$.}
\label{ff1p}
\end{figure}

The potential (\ref{VVpot}) has different forms for various values of dilaton coupling. 
Fig.\ref{ff1p} is standing the potential as the variable $R$ for $a^2=0,1/3$ and $1$. 
We can find that there is the potential barrier in the case of $a^2=0$. 
When a potential barrier exists, the incoming particles are scattered or captured. 
From the equation (\ref{varR}), the limit of $r\to 0$ is corresponding to the limit of $R\to -\infty$. 
If the incoming particles through the potential barrier can approach 
to $-\infty$ infinitely, then two black holes can approach to each other. 
In the case of $a^2=1$, the potential has the infinity wall at $r=0$ ($R=0$). 
Therefore the incoming particles are scattered away. 
In the case of $a^2=1/3$, the potential has the barrier for $l=0$.
For $l=0$, the incoming particles are scattered or coalesced. 
For $l\neq0$, there is the finite wall. 
In the case of $l\neq0$, the incoming particles which have the higher energy than a potential barrier are scattered or coalesced, and the low energy particles are always scattered away. 
%%%%%%%%%%%%%%%%%%%%%%%%%%%%%%%%%%%%%%%%%%%%%%%%%%%%%%%%%%%%%%%%%%%%%%%%%%%

\section{The asymptotic solutions}
In this section, we consider the asymptotic states of two black holes system. 
Because the Schr\"{o}dinger equation depends on the dilaton coupling, we 
consider the case of each values of $a$. 
In the case $a^2=0$, the black holes are corresponding to Reisner-Nordstr{\"o}m Black holes discussed in \cite{fe, TrFe}. 
We consider the asymptotic solutions in the case of $a^2=1$ and $a^2=1/3$. 
%%%%%%%%%%%%%%%%%%%%%%%%%%%%%%%%%%%%%%%%%%%%%%%%%%%%%%%%%%%%%%%%%%%%%
\subsection{The case of $a^2=1$}
In the case of $a^2=1$, the moduli space metric (\ref{modmet2}) is 
given by 
\begin{equation}
\gamma=1+\frac{2M}{r}, \\
\end{equation}
and the variable $R$ and potential $V$ are written by 
\begin{eqnarray}
R&=&\sqrt{r(r+2M)}+2M \ln\left(\sqrt{\frac{r}{2M}}+\sqrt{\frac{2M+r}{2M}}\right), \label{R1} \\
V&=&\frac{M}{(2M+r)^3}+\frac{l(l+1)}{r(2M+r)}, \label{V1}
\end{eqnarray}
where $0\leq R<\infty$. 

In the limit of $r/M\to0$ ($R/M \to 0$), (\ref{R1}) and (\ref{V1}) are given by following form: 
\begin{eqnarray}
R&=&\sqrt{2Mr},\\
V&=&\frac{1}{8M^2+6R^2}+\frac{l(l+1)}{R^2}. \label{potvapp1}
\end{eqnarray}
When $r/M\to \infty$($R \gg M$), 
\begin{eqnarray}
R&=&r+2M \ln(2 \sqrt{\frac{r}{2M}}), \\
V&=&\frac{M}{R^3}+\frac{l(l+1)}{R^2}.
\end{eqnarray}

As seen in sec III, the potential has the infinite wall at $R=0$. 
Then, the incoming particles are always scattered away. 
The states will be the sum of incoming wave $e^{-iqR}$ and outgoing wave $e^{+iqR}$. 
For $r/M\to0$ ($R/M \to 0$), the states are given by 
\begin{equation}
\chi_{ql}=\sqrt{R}\left[ a(q) J_n(ER) + b(q) J_{-n}(ER) \right],
\label{asymnpsol1}
\end{equation}
where 
\begin{eqnarray}
n&=&l+1/2,\\
E&=&\sqrt{q^2-1/8M^2}.
\end{eqnarray}

And the states for $r/M\to \infty$($R \gg M$) are written by
\begin{equation}
\chi_{ql} = \sqrt{R} \left[ a(q) J_n (qR) + b(q) J_{-n}(qR) \right].
\label{asymnpsol2}
\end{equation}

The Hamiltonian (\ref{4scheq2}) includes an inverse square interaction. 
In the limit of $r/M\to0$ ($R/M\to0$), the behaviors of potential are dominated by the inverse square term. 
The quantum systems with inverse square interaction are discussed in \cite{GSV, Basu-Mallick:2002ni}. 
When the coefficient of inverse square interaction is represented by $C$, 
the typical Hamiltonian $H$ is 
\begin{equation}
H=-\frac{d^2}{dR^2}+\frac{C}{R^2}. 
\end{equation}
This form of Hamiltonian is an unbounded differential operator defined in $R^+ \equiv[0,\infty]$.
$H$ is a symmetric operator on the domain 
$D(H)\equiv \{  \phi \in L^2[R^+,du],\ \phi(0)=\phi'(0)=0 \}$. 
It is known that $H$ for $C\geq 3/4$ is self-adjoint on the domain $D(H)$. 
For $-1/4\leq C <3/4$, $H$ is not self-adjoint on the domain $D(H)$ but 
admits self-adjoint extensions, where 
the self-adjoint extensions are labeled by a $U(1)$ parameter $e^{iz}$. 
Each value of parameter $z$ defines the domain $D_{z}(H)$. 
The operator $H$ is self-adjoint in the domain $D_{z}(H)$ which contains 
all the vector in $D(H)$ and the vectors of 
form $\phi_{+}(u)+e^{iz} \phi_{-}(u)$, where
\begin{eqnarray}
\phi_{+}=\sqrt{R} H_{\nu}^{(1)}(R e^{i\pi /4}), \\
\phi_{-}=\sqrt{R} H_{\nu}^{(2)}(R e^{-i\pi /4}),
\end{eqnarray}
where $\nu=\sqrt{1/4+C}$, and $H_{\nu}^{(1,2)}$ are Hankel functions. If $C=l(l+1)$, then $\nu=n=l+1/2$. 

In the only case of $l=0$, the potential (\ref{potvapp1}) satisfies the condition of $-1/4\leq C <3/4$. 
Then, the Hamiltonian admits self-adjoint extentions. 
For $r\to 0$, 
\begin{equation}
\phi_+ (R)+ e^{iz} \phi_- (R) 
\approx i 
\left[ \frac{R}{\sqrt{2}} \frac{ e^{-i\frac{3\pi}{8}}-
e^{i(z+\frac{3\pi}{8})}}{\Gamma(\frac{3}{2})} 
+\frac{1}{\sqrt{2}}\frac{e^{i(z+\frac{\pi}{8})}-e^{-i\frac{\pi}{8}}}{\Gamma(\frac{1}{2})} \right]. \label{extlim}
\end{equation}
The asymptotic states (\ref{asymnpsol1}) in the limit of $r\to 0$ are given by following form:   
\begin{equation}
\chi_{q0}\approx \frac{R}{\sqrt{2}}\frac{a(q) \sqrt{E}}{\Gamma(\frac{3}{2})}
+\sqrt{2}\frac{b(q)}{\sqrt{E}\Gamma(\frac{1}{2})} \label{chilim}.
\end{equation}
If $\chi_{q0}$ are the vectors on the domain $D_z (H)$, then the coefficient of $R$ in equation (\ref{extlim}) and (\ref{chilim}) must match each other. 
From these coefficients, we get 
\begin{equation}
\frac{a(q)}{b(q)}=-\frac{1}{E} \frac{\sin (\frac{z}{2}+\frac{3\pi}{8})}{\sin (\frac{z}{2}+\frac{\pi}{8})} \label{coef}.
\end{equation}
The asymptotic expansion of equation (\ref{asymnpsol2}) is represented by the following form: 
\begin{eqnarray}
\chi_{ql}\approx& &\frac{1}{\sqrt{2\pi q}}e^{iqR}\left[a(q)e^{-i(\nu+\frac{1}{2})\frac{\pi}{2}}
+b(q) e^{i(\nu-\frac{1}{2})\frac{\pi}{2}} \right] \nonumber \\
& &\hspace{5mm}+\frac{1}{\sqrt{2\pi q}}e^{-iqR}\left[a(q)e^{i(\nu+\frac{1}{2})\frac{\pi}{2}}
+b(q)e^{-i(\nu-\frac{1}{2})\frac{\pi}{2}} \right]. \label{coefas}
\end{eqnarray}
The S-matrix is obtained by dividing the coefficient of outgoing wave 
by that of the incoming wave. 
Using (\ref{coef}) and (\ref{coefas}), we get the S-matrix for $l=0$. 
\begin{equation}
S(q)\equiv e^{2i\delta(q)}=\frac{-iE^{-\frac{1}{2}} \sin(\frac{z}{2}+\frac{3\pi}{8})
-E^{\frac{1}{2}} \sin(\frac{z}{2}+\frac{\pi}{8})}
{iE^{-\frac{1}{2}} \sin(\frac{z}{2}+\frac{3\pi}{8})
-E^{\frac{1}{2}} \sin(\frac{z}{2}+\frac{\pi}{8})}, \label{smat}
\end{equation}
where 
\begin{eqnarray}
E=\sqrt{q^2-\frac{1}{8M^2}}, \\
\end{eqnarray}
and $z$ is the self-adjoint extention parameter. 

\begin{figure}[htpb]
\centering
\includegraphics{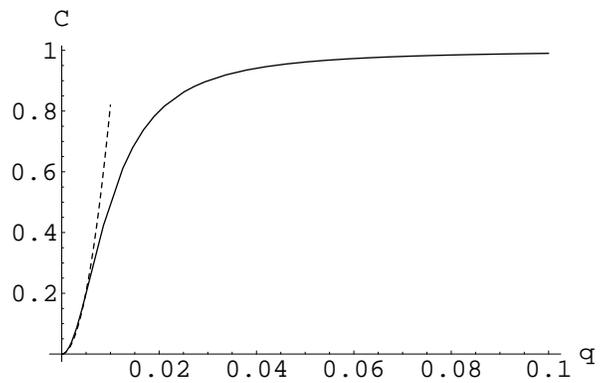}
\vspace*{10mm}
\caption{The capture coefficient for $a^2=1/3$ and $l=0$. The dashed line is given by the asymptotic states in the near-coincident black holes limit $R\to-\infty$. The solid line is calculated with the $\delta$ function potential.}
\label{ff2p}
\end{figure}

For $l\neq0$, the coefficient of the inverse square term satisfies $C>3/4$. 
The Hamiltonian is self-adjoint in the domain $D(H)$. 
In this case, the scattering states (\ref{asymnpsol2}) for 
the region of $R\gg M$ become 
\begin{equation}
\chi_{ql}=a(q)\sqrt{R}J_{n}(qR). \label{selfadjoint}
\end{equation}
From the asymptotic form of (\ref{selfadjoint}), 
the S-matrix is written by the following form: 
\begin{equation}
S=\exp[-i \pi(n+\frac{1}{2})]=(-1)^{l+1}. 
\end{equation}
The partial cross section is
\begin{equation}
\sigma_l=\frac{4\pi}{q^2}(2l+1)\sin^2\delta_l=\frac{4\pi}{q^2}(2l+1). \label{ew45}
\end{equation}
In Fig.\ref{ff2p}, 
the partial cross sections for $l=10$ and $l=20$ are plotted. 
The solid lines in Fig.\ref{ff2p} are plotted by (\ref{ew45}) and the points are the results calculated with WKB approximation (see \cite{Sakamoto:2003ju}). 
These results have the same behavior. 

The asymptotic states (\ref{asymnpsol1}) show the quantum states in the moduli space constructed with the near-coincident black holes. 
The bound states in the moduli space for the limit of near-coincident black holes are discussed in \cite{Sakamoto:2002ci}. 
The states (\ref{asymnpsol1}) include the bound states obtained in \cite{Sakamoto:2002ci}. 
%%%%%%%%%%%%%%%%%%%%%%%%%%%%%%%%%

%%%%%%%%%%%%%%%%%%%%%%%%%%%%%%%%%%%%%%%%%%%%%%%%%%%%%%%%%%%%%%%%%%%%%%%%%%%

\subsection{The case of $a^2=1/3$}
In the case of $a^2=1/3$, the moduli space metric (\ref{modmet2}) becomes
\begin{equation}
\gamma=(1+\frac{4M}{3r})^2,
\end{equation}
and the variable $R$ and potential $V$ are given by
\begin{eqnarray}
R&=&r+\frac{4M}{3}\ln (\frac{3r}{4M}),\label{R13} \\
V&=&\frac{108Mr}{(4M+3r)^4}+
\frac{9 l(l+1)}{(4M+3r)^2},\label{V13}
\end{eqnarray}
where $-\infty<R<\infty$. 
The potential has the barrier for $l=0$ and the finite wall for $l\neq0$. 

When $r/M\to0$ ($R \ll -M$), (\ref{R13}) and (\ref{V13}) are given by the following form: 
\begin{eqnarray}
R&=&\frac{4M}{3}\ln(\frac{3r}{4M}),\\
V&=&\frac{9}{16M^2}e^{\frac{3R}{4M}}+\frac{9l(l+1)}{16M^2}, 
\end{eqnarray}

and when $r/M \to \infty$ ($R \gg M$), 
\begin{eqnarray}
R&=&r, \\
V&=&\frac{l(l+1)}{R^2}. 
\end{eqnarray}

As seen in sec III, for $l=0$ or $l\neq 0$ and the higher energy than finite wall of potential, the incoming particles are scattered or coalesced. 
When $R\to -\infty$, the states are described by the only left moving wave. 
In the limit  $R\to +\infty$, the states will be the sum of incoming wave $e^{-iqR}$ and outgoing wave $e^{+iqR}$. 
The low energy particles for $l\neq0$ are always scattered away. 
Then the states will be the sum of incoming wave and outgoing wave in the limit of $R\to +\infty$. 

First, we consider the case of black holes coalesce. 
The asymptotic solutions of (\ref{4scheq2}) for $r/M\to0$ are written by
\begin{eqnarray}
\chi_{ql}= C_{ql}\  H_\mu^{(-)} \left( i 2 e^{\frac{3R}{8M}} \right), \label{state13a}\\
\mu=\frac{2}{3}\sqrt{9l(l+1)-16q^2 M^2}. 
\end{eqnarray}
For $r/M \to \infty$, the asymptotic solutions are given by the following form: 
\begin{equation}
\chi_{ql}= qR\left[ (-i)^{l+1} h_l^{(-)} (qR)+S_{ql}h_l^{(+)}(qR) \right]. \label{state13b}
\end{equation}
%In the case of $a^2=1/3$, as described in sec III, the potential has the 
%barrier for $l=0$. Then the incoming particles are coalesced. 
We discuss the low energy states. 
In the limit of $qM\to0$, the state (\ref{state13b}) are 
\begin{equation}
\chi_{ql} \approx \frac{(S_{ql}+(-i)^{l+1})(qR)^{l+1}\sqrt{\pi} }
{2^{l+1}\Gamma(l+\frac{3}{2})}
-i\frac{ 2^l (S_{ql}-(-i)^{l+1})\Gamma(l+\frac{1}{2})}
{(qM)^l(R/M)^l\sqrt{\pi} }, \label{state13c}
\end{equation}
in the region $M\ll R\ll\frac{1}{q}$. 
The states (\ref{state13a}) are represented by 
\begin{equation}
\chi_{ql}\approx
C_{ql}(\frac{1}{i\pi\exp(\frac{3R}{8M})})^{1/2}
\exp \left(2\exp(\frac{3R}{8M})+i(\frac{\pi}{4} + \frac{\pi}{3}\sqrt{9l(l+1)-16q^2 M^2})\right), \label{state13d}
\end{equation}
in the region $-\frac{1}{q}\ll R\ll-M$. 
As the differential equation (\ref{4scheq2}) is roughly independent of $q$ in the region $-\frac{1}{q}\ll R \ll\frac{1}{q}$, the dependence on $q$ 
for the states (\ref{state13a}) and (\ref{state13b}) must be same. 
In the limit $qM\to 0$, we take $S_{ql}\approx(-i)^{l+1}+(qM)^{2l+2}s_{ql}$, where $|s_{ql}|\leq1$. This is self-consistent (see \cite{TrFe}). 
Using (\ref{state13c}) and (\ref{state13d}), 
\begin{equation}
|C_{ql}|^2=\frac{\pi^2 (qM)^{(2l+2)}|\kappa_l|^2}{2^{2l}(\Gamma(l+3/2))^2}
\exp\left[\frac{3}{8}-4e^{\frac{3}{8}} \right] \label{capcoe}
\end{equation}
where $\kappa_l$ is an unknown parameter.
In the case of $l\neq0$, the capture coefficient (\ref{capcoe}) approaches to zero quickly in the low energy limit. 
This means that the low energy incoming particles for $l\neq0$ are scattered away, as seen in the discussion of the potential form, and the significant capture occurs for $l=0$. 
So, two black holes cannot coalesce when the states of two-black hole system 
have the low energy and $l\neq0$. 
For $l=0$, the capture coefficient is written by
\begin{equation}
|C_{q0}|^2= \frac{ \pi^2 (qM)^2|\kappa_0|^2}{ (\Gamma(\frac{3}{2}))^2} \exp\left[\frac{3}{8}-4e^{\frac{3}{8}} \right] . 
\label{capq0}
\end{equation}

In Fig.\ref{ff2p}, the capture coefficient for $l=0$ is shown. 
The dashed line is given by (\ref{capq0}) for a suitable value of parameter $\kappa_0$. 
The solid line represents the capture coefficient obtained from 
the approximate potential with a $\delta$ function (see \cite{fe}). 
In the low energy limit, these capture coefficients become like $|C_{q0}|^2\propto (qM)^2$ and correspond to each other. 
As same as the case of two black holes with no dilaton coupling \cite{fe}, 
the behavior is different from the classical particle. 
Classically, the two black holes that approach each other with zero angular momentum always coalesce. 
In the quantum mechanics, we can expect that the black holes for low energy and $l=0$ scatter away. 

\begin{figure}[htpb]
\centering

\includegraphics{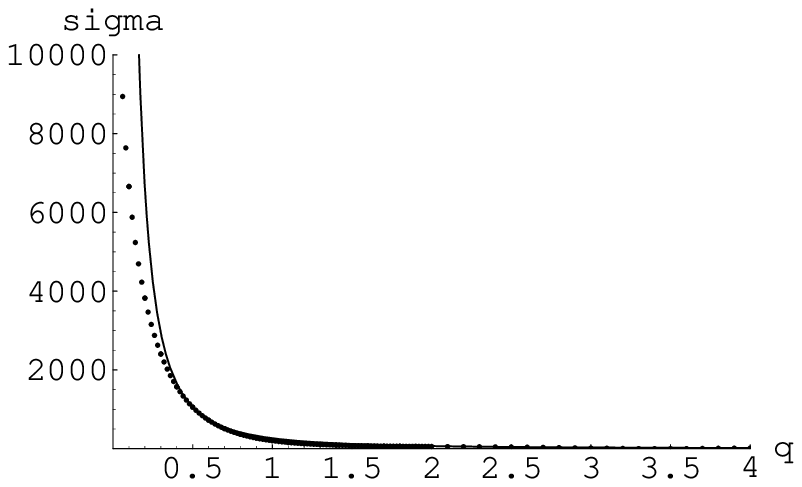}\\
\vspace*{-3mm}
A

\vspace*{10mm}
\includegraphics{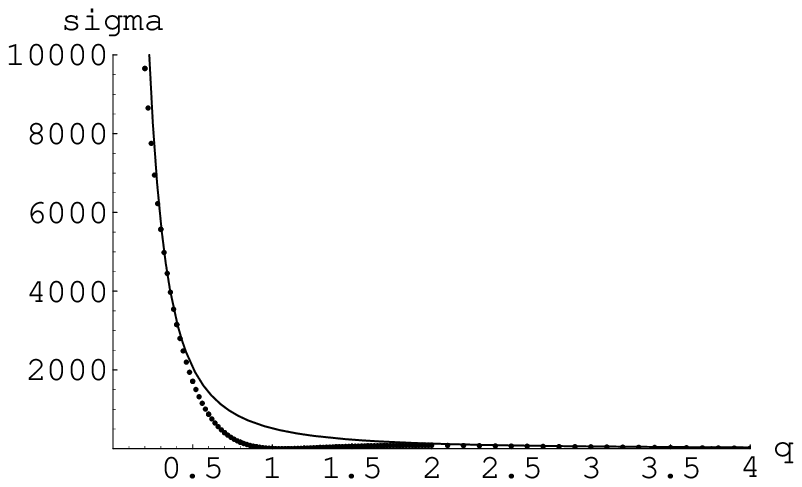}\\
\vspace*{-3mm}
B

\vspace*{10mm}
\caption{The partial cross section in the case of $a^2=1/3$. A for $l=10$ and B for $l=20$. The solid lines are given by the asymptotic states and the points are the results calculated with WKB approximation. }
\label{ff4p}
\end{figure}

Next, we consider the states for $l\neq0$ and the lower energy than finite wall of potential. 
In this case, the incoming particles are scattered away. 
We can consider the same way as the case of $a^2=1$. 
The potential becomes an inverse square interaction for $r/M\to\infty$. 
The Hamiltonian is self-adjoint in the domain $D(H)$. 
In this case, the solutions are written by following form:
\begin{equation}
\chi=a(q)\sqrt{R}J_{n}(qR). 
\end{equation}
The S-matrix and the partial cross section are also given the same way as the case of $a^2=1$. 
In Fig.\ref{ff4p}, 
the partial cross sections obtained from the asymptotic states and the results calculated with WKB approximation (see \cite{Sakamoto:2003ju})
 are plotted. 
These results have the same behavior.

%%%%%%%%%%%%%%%%%%%%%%%%%%%%%%%%%%%%%%%%%%%%%%%%%%%%%%%%%%%%%%%%%%%%%%%%%%%
\section{conclusion}
In this paper we discussed the quantum states in the moduli space which constructed with two maximally charged black holes with dilaton. 
Using the moduli space metric, we calculated the Schr\"{o}dinger equation in the moduli space. 
In the limit of the near-coincident black holes or the widely separated black holes, we obtained the asymptotic solutions. 
Using these solutions, we discussed the black hole-black hole scattering. 
In the case of $a^2=1$, the incoming particles are always scattered away. 
The behaviors of partial cross section for $a^2=1$ were similar to the results of WKB approximation discussed in \cite{Sakamoto:2003ju}. 
In the case of $a^2=1/3$, we studied the captured coefficient. 
As same as the results in \cite{fe}, 
the quantum effects for the dilaton black holes were seen in the captured coefficient for $l=0$.
For $l\neq0$, the incoming particles with low energy are always scattered away. 
The behaviors of partial cross section obtained from asymptotic states were corresponding to the results with the WKB approximation. 

The geometries of moduli space consisting of maximally charged black holes with dilaton coupling were discussed in \cite{shir}. 
The moduli spaces have various geometries for arbitrary values of dilaton coupling. 
The quantum effects for difference of these geometries cannot find in this paper. The quantum effects were studied in \cite{Sakamoto:2003ju} with semi-classical approximation, where we found the resonance states in the scattering process for $1/3<a^2<1$. 
In this paper, we discussed the quantum states for $a^2=1/3$ and $1$, because the moduli space metric becomes a simple form. 
For the other values of dilaton coupling, such as $1/3<a^2<1$, we would like to discuss the quantum states. 
In other further studies, we should consider the quantum states more precisely. 
In this paper, we considered the first order of the variable $R$ and potential $V$ in the limit. 
Considering the higher order, we expect that more precise states are obtained.

%%%%%%%%%%%%%%%%%%%%%%%%%%%%%%%%%

%%%%%%%%%%%%%%%%%%%%%%%%%%%%%%%%%
%%%%%%%%%%%%%%%%%%%%%%%%%%%%%%%%%%

%%%%%%%%%%%%%%%%%%%%%%%%%%%%%%%%%
\begin{figure}[htpb]
\centering

\includegraphics{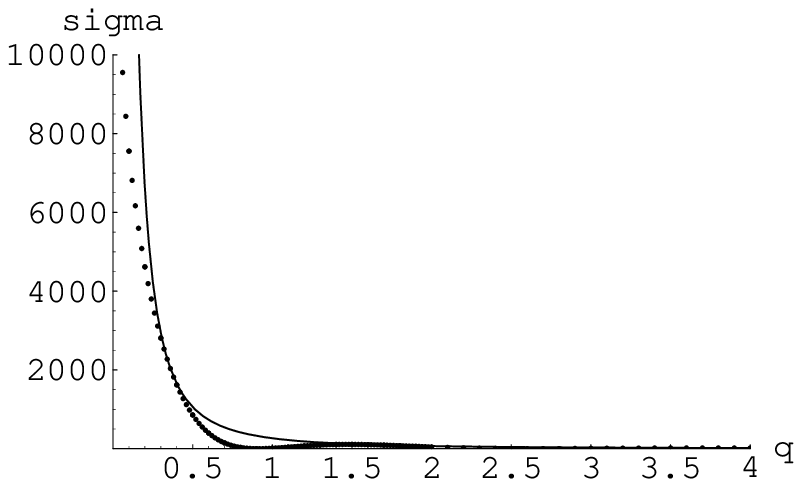}\\
\vspace*{-3mm}
A

\vspace*{10mm}
\includegraphics{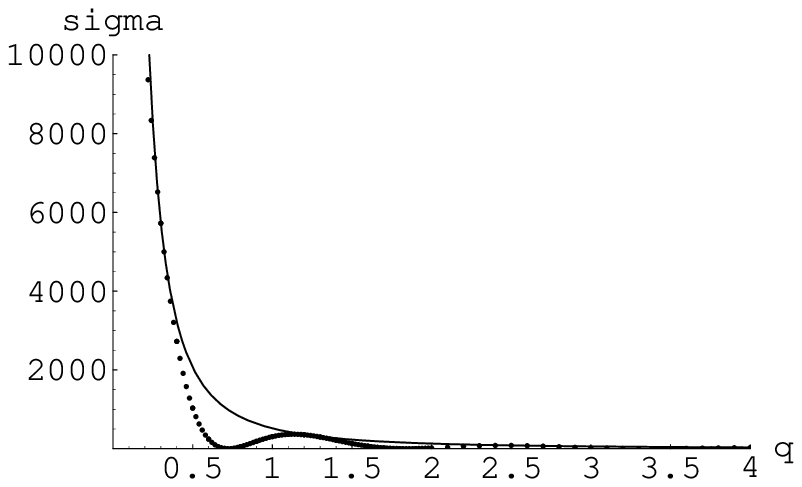}\\ 
\vspace*{-3mm}
B

\vspace*{10mm}
\caption{The partial cross section in the case of $a^2=1$. A for $l=10$ and B for $l=20$. The solid lines are given by the asymptotic states and the points are the results calculated with WKB approximation.}
\label{ff5p}
\end{figure}
%%%%%%%%%%%%%%%%%%%%%%%%%%%%%%%%%
%%%%%%%%%%%%%%%%%%%%%%%%%%%%%%%%%
%%%%%%%%%%%%%%%%%%%%%%%%%%%%%%%%%%

%%%%%%%%%%%%%%%%%%%%%%%%%%%%%%%%%
%%%%%%%%%%%%%%%%%%%%%%%%%%%%%%%%%


\begin{thebibliography}{999}


\bibitem{Papapetrou:1947ib}
A.~Papapetrou,
Proc.\ Roy.\ Irish Acad.\ (Sect.\ A)A {\bf 51}, 191 (1947).

\bibitem{Majumdar:eu}
S.~D.~Majumdar,
Phys.\ Rev.\  {\bf 72}, 390 (1947).


\bibitem{IsraelKhan}
W.~Israel and K.~A.~Khan,
Nuovo\ Cim.\  {\bf 33},\ 331

\bibitem{KS3}
K.~Shiraishi,
Int.\ J.\ Mod.\ Phys.\ D {\bf 2}, 59 (1993).

\bibitem{Myers:rx}
R.~C.~Myers,
Phys.\ Rev.\ D {\bf 35}, 455 (1987).

\bibitem{Kastor:1992nn}
D.~Kastor and J.~H.~Traschen,
D {\bf 47}, 5370 (1993)

\bibitem{Emparan:1999au}
R.~Emparan,
Phys.\ Rev.\ D {\bf 61}, 104009 (2000).

\bibitem{Emparan:2001bb}
R.~Emparan and E.~Teo,
%``Macroscopic and microscopic description of black diholes,''
Nucl.\ Phys.\ B {\bf 610}, 190 (2001)

\bibitem{Jones:2004rg}
G.~Jones, A.~Maloney and A.~Strominger,
Phys.\ Rev.\ D {\bf 69}, 126008 (2004)

\bibitem{Gibbons}
G.~W.~Gibbons and P.~J.~Ruback,
Phys.\ Rev.\ Lett.\  {\bf 57}, 1492 (1986).

\bibitem{fe} 
R.~C.~Ferrell and D.~M.~Eardley,
Phys.\ Rev.\ Lett.\  {\bf 59} 1617 (1987).

\bibitem{TrFe} 
J.~Traschen and R.~Ferrell,
Phys.\ Rev.\ D {\bf 45}, 2628 (1992).

\bibitem{shir} 
K.~Shiraishi,
Nucl.\ Phys.\ B {\bf 402}, 399 (1993).

\bibitem{Shiraishi:1992ct}
K.~Shiraishi,
Int.\ J.\ Mod.\ Phys.\ D {\bf 2}, 59 (1993).

\bibitem{Sakamoto:2003ju}
K.~Sakamoto and K.~Shiraishi,
Phys.\ Rev.\ D {\bf 68}, 025019 (2003)

\bibitem{as2}
A.~Maloney, M.~Spradlin and A.~Strominger,
JHEP {\bf 0204}, 003 (2002).

\bibitem{InPl} 
L.~Infeld and J.~Pleba\'nski, {\it Motion and Relativity}, 
(Pergamon Press, 1960).

\bibitem{GSV} 
T. R. Govindarajan, V. Suneeta and S. Vaidya, Nucl. Phys. {\bf B583}, 291 (2000).

\bibitem{Basu-Mallick:2002ni}
B.~Basu-Mallick, P.~K.~Ghosh and K.~S.~Gupta,
Nucl.\ Phys.\ B {\bf 659}, 437 (2003)

\bibitem{Sakamoto:2002ci}
K.~Sakamoto and K.~Shiraishi,
Phys.\ Rev.\ D {\bf 66}, 024004 (2002).

\end{thebibliography}
\end{document}